\documentclass{moriond}

\def\Journal#1#2#3#4{{#1} {\bf #2}, #3 (#4).}

\def\JHEP{\em JHEP}

\def\NIMA{{\em Nucl. Instrum. Methods} A}

\def\PLB{{\em Phys. Lett.}  B}
\def\PRL{\em Phys. Rev. Lett.}
\def\PRD{{\em Phys. Rev.} D}
\def\PRP{\em Phys. Rept.}

\def\EPJ{{\em Eur. Phys. J.} C}

\def\be{\begin{equation}}
\def\ee{\end{equation}}
\def\bea{\begin{eqnarray}}
\def\eea{\end{eqnarray}}

\usepackage{relsize}
\def\babar{\mbox{\slshape B\kern-0.1em{\smaller A}\kern-0.1em
    B\kern-0.1em{\smaller A\kern-0.2em R}}}

\def\pho{\textsc{Phokhara}}

\def\GeVM{~${\rm GeV}/c^2$}
\def\MeVM{~${\rm MeV}/c^2$}
\def\GeVP{~${\rm GeV}/c$}
\def\fb{~${\rm fb}^{-1}$}

\newcommand{\mumug}{$\mu\mu\gamma$}
\newcommand{\pipig}{$\pi\pi\gamma$}
\newcommand{\KKg}{$KK\gamma$}
\newcommand{\eeg}{$ee\gamma$}
\newcommand{\mmumu}{$m_{\mu\mu}$}
\newcommand{\mpipi}{$m_{\pi\pi}$}

\newcommand{\eepipig}{$e^+e^- \rightarrow \pi^+\pi^-(\gamma)$}
\newcommand{\eemumug}{$e^+e^- \rightarrow \mu^+\mu^-(\gamma)$}

\usepackage{subfig}
\usepackage{lineno}

\newcommand{\Photo}{}

\begin{document}
\title{New precise measurement of the \eepipig\ cross section with \babar}

\author{ L\'eonard Polat, on behalf of the \babar\ collaboration }

\address{LPNHE, 4 place Jussieu, 75252 Paris Cedex 05, France\\
IJCLab, 15 rue Georges Clémenceau, 91405 Orsay, France\\
\vspace{0.4cm}
{\small {\normalfont Presented at the 32nd International Symposium on Lepton Photon Interactions at High Energies, Madison, Wisconsin, USA, August 25-29, 2025.}}}

\maketitle\abstracts{
The \babar\ experiment participates to the global endeavor for a precise prediction of the anomalous magnetic moment of the muon by evaluating the contribution from hadronic vacuum polarization, in particular through cross section measurements of hadronic final states from $e^+e^-$ collisions. After a first measurement in 2009 of the largest input that comes from the \eepipig\ cross section, we present preliminary results from a new study on 460\fb\ of \babar\ data, involving a blind and independent procedure. The results of the two analyses are shown to be consistent.
}

\section{Introduction}

The anomalous magnetic moment of the muon $a_\mu$ is sensitive to hadronic vacuum polarization (HVP), which is the dominant source of uncertainty on its predicted value~\cite{wp25}. It is therefore of great interest for physicists to improve the precision on the HVP contribution to $a_\mu$, in the search for potential tensions with direct measurements.

\begin{figure}[ht]
\centering
\subfloat[]{\includegraphics[width=0.48\columnwidth,trim={1cm 0 1cm 1cm}]{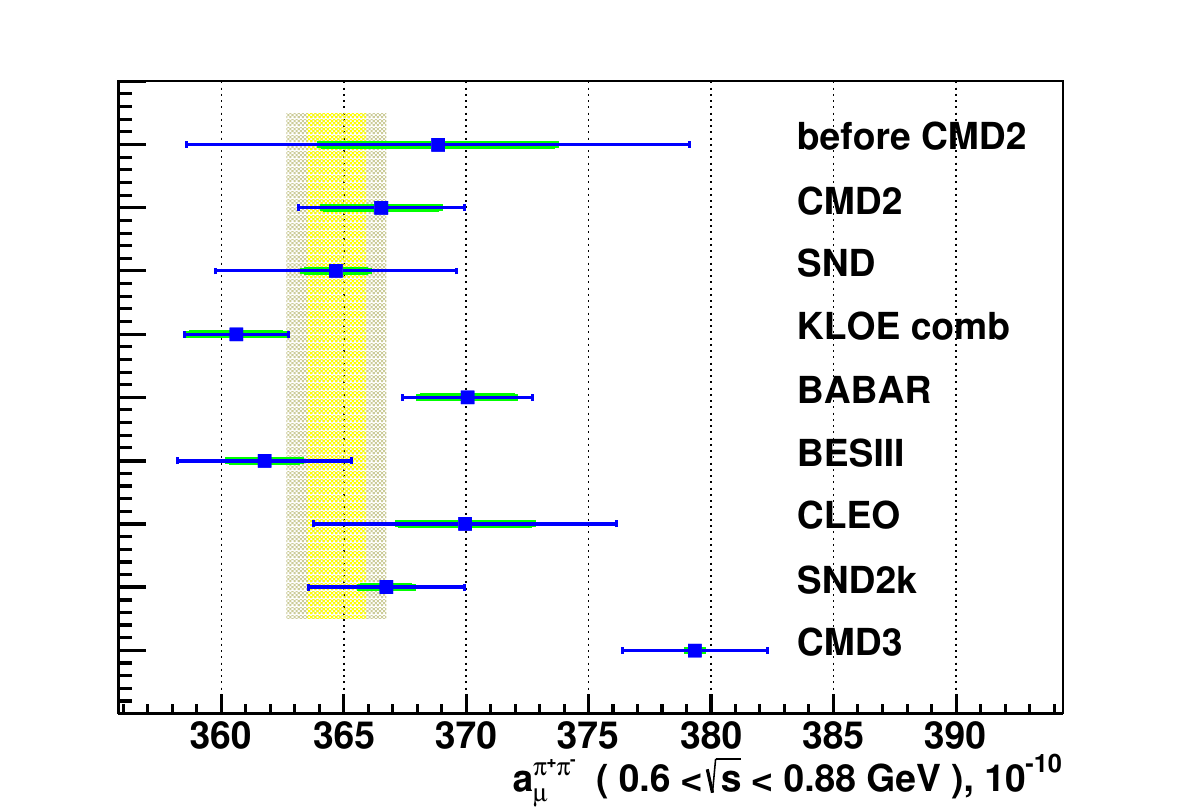}}
\hspace{0.4cm}
\subfloat[]{\includegraphics[width=0.41\columnwidth]{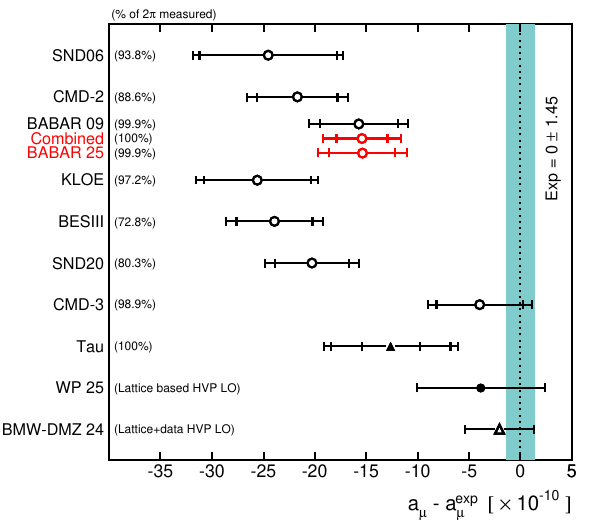}}
\vspace{-0.2cm}
\caption{
The $\pi^+\pi^-$ contribution to $a_\mu$ in the energy range $0.6 < \sqrt{s} < 0.88$ GeV obtained from multiple experiments~\protect\cite{cmd} (a) and comparison of dispersive predictions (preliminary new results in red) with the direct experimental measurement (vertical band) and lattice-based results (b).}
\label{fig:amu}
\end{figure}

This contribution can be obtained through a dispersion integral by measuring the cross sections of $e^+e^- \rightarrow \mathrm{hadrons}$ processes. The largest input comes from $e^+e^- \rightarrow \pi^+\pi^-$ and has been measured by many experiments using the initial state radiation (ISR) method (among the most recent ones \babar~\cite{babar1,babar2}, KLOE~\cite{kloe}, CLEO-c~\cite{cleo}, BESIII~\cite{bes}) or direct energy scans (SND~\cite{snd}, CMD-3~\cite{cmd}), their results summarized in Figure~\ref{fig:amu} (a) in a narrow energy range at the $\rho$ meson peak. As of today, some of these dispersive predictions are in tension with each other, especially KLOE and CMD-3 up to more than $5\sigma$ at the $\rho$ energy~\cite{dhlmz}. Tensions exist as well with the direct measurements of $a_\mu$ and the calculation from lattice QCD, shown in Figure~\ref{fig:amu}~(b). Thus, it is necessary to conduct further studies to solve these discrepancies.

The last measurement of the \eepipig\ cross section at \babar\ was published in 2009. We present preliminary results on a new measurement that involves an independent channel separation method and twice as much data statistics collected by the experiment.

\vspace{-0.5cm}

\section{The \texorpdfstring{\babar}{BaBar} data and simulation samples}
\label{sec:babar}

\babar~\cite{exp} is an experiment that operated from 1999 to 2008 at the SLAC National Accelerator Laboratory (USA). It exploited asymmetric collisions of electrons and positrons injected in the storage rings of the PEP-II facility, with respective energies of 9 and 3 GeV and total center-of-mass (c.m.) energy $\sqrt{s}=10.58$\GeVM, at the $\Upsilon(4S)$ meson resonance. The experiment collected 424.2\fb\ of data at this resonance and 43.9\fb\ off-resonance~\cite{lumi}.

In addition to the collected data, Monte Carlo (MC) simulation samples are generated with \pho~\cite{pho} for signal events, namely the $\pi^+\pi^-(\gamma)\gamma_{\mathrm{ISR}}$ and $\mu^+\mu^-(\gamma)\gamma_{\mathrm{ISR}}$ final states, where $\gamma_{\mathrm{ISR}}$ stands for the main ISR photon.
Background samples that simulate the processes $e^+e^- \rightarrow q\bar{q}~(q=u,d,s,c),~\tau^+\tau^-,~X\gamma_{\mathrm{ISR}}~(X=K^+K^-,~n\pi/K+m\pi^0,...)$ are also generated.

\section{Differences between the 2009 and 2025 cross section measurements at \texorpdfstring{\babar}{BaBar}}

The last \babar\ analysis measured the \eepipig\ cross section as a function of the reduced energy $\sqrt{s^\prime}=m_{\pi^+\pi^-(\gamma)}$ that includes additional FSR photons, using 232\fb\ of data on and off the $\Upsilon(4S)$ resonance. To cancel out common systematic effects, the measured $\pi^+\pi^-(\gamma)$ mass spectrum is divided by the $\mu^+\mu^-(\gamma)$ spectrum, equivalent to the ratio of each final state's bare cross section. The separation between charged pion and muon tracks was based on particle identification (PID), which required the selection $p>1$\GeVP\ on each track momentum to make the muon identification more reliable. In the end, the total relative systematic uncertainty on the cross section around the $\rho$ peak was 0.5\%, dominated by PID.

In this new \babar\ analysis, around 460\fb\ of data are studied, while PID requirements on the tracks are removed. An angular fit is considered as a new method~\cite{method} to separate the main signal and background processes, based on the absolute value of cosine of the angle between the negative charge track and the ISR photon in the 2-track c.m. frame, $|\cos\theta^*|$. To improve the distinction between the dipion and dimuon distribution shapes, the $p>1$\GeVP\ selection on the track momentum in the 2009 analysis is replaced by a looser selection on the transverse momentum $p_\mathrm{T}>0.1$\GeVP, increasing the statistics at the same time.

\section{Methodology of the analysis and results}

All events go through two kinematic fits based on additional radiation at next-to-leading order (NLO), depending on whether an NLO photon is emitted at large (LA) or small (SA) angle from the beams. These kinematic fits are performed twice, assuming either the muon or pion mass hypotheses for the charged tracks and therefore allow to get the masses $m_{XX}$ and angular distributions $|\cos\theta^*_{X}|$ in both bases ($X=\mu$ or $\pi$). Most of background processes in data are separated from signal thanks to the $\chi^2$ values of the fits, optimizing with boosted decision trees (BDTs) a two-dimensional $\chi^2_{\mathrm{LA}}$ vs $\chi^2_{\mathrm{SA}}$ (2D-$\chi^2$) selection that retains more than 98\% of signal.

Minor remaining background processes are subtracted from data according to simulation, leaving only the final states \pipig, \mumug, \KKg\ and \eeg \footnote{The $X^+X^-(\gamma)$ notation is reduced to $XX\gamma$ for simplicity.}. $|\cos\theta^*|$ distributions in data are fitted with templates of these four channels: the first three ones are obtained from corrected MC samples, while the last background template is extracted from data, enriched with \eeg\ events through cut-based and BDT selections, as there is no reliable simulation for this process.

\begin{figure}[ht]
\centering
\subfloat[]{\includegraphics[width=0.48\columnwidth]{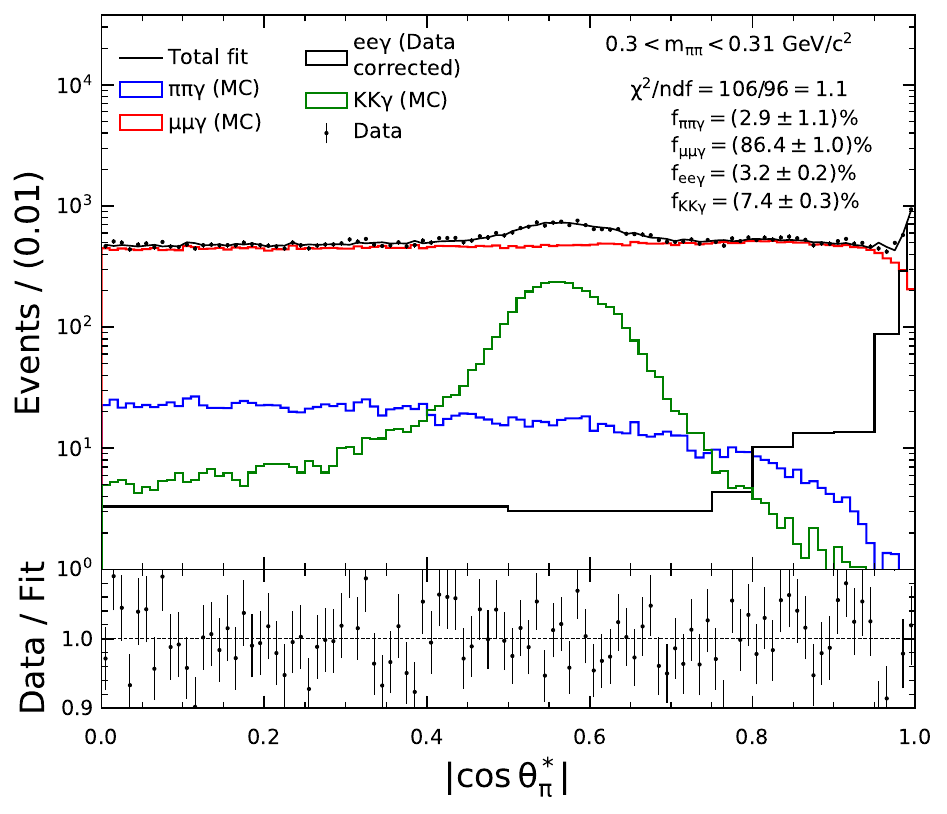}}
\hspace{0.4cm}
\subfloat[]{\includegraphics[width=0.48\columnwidth]{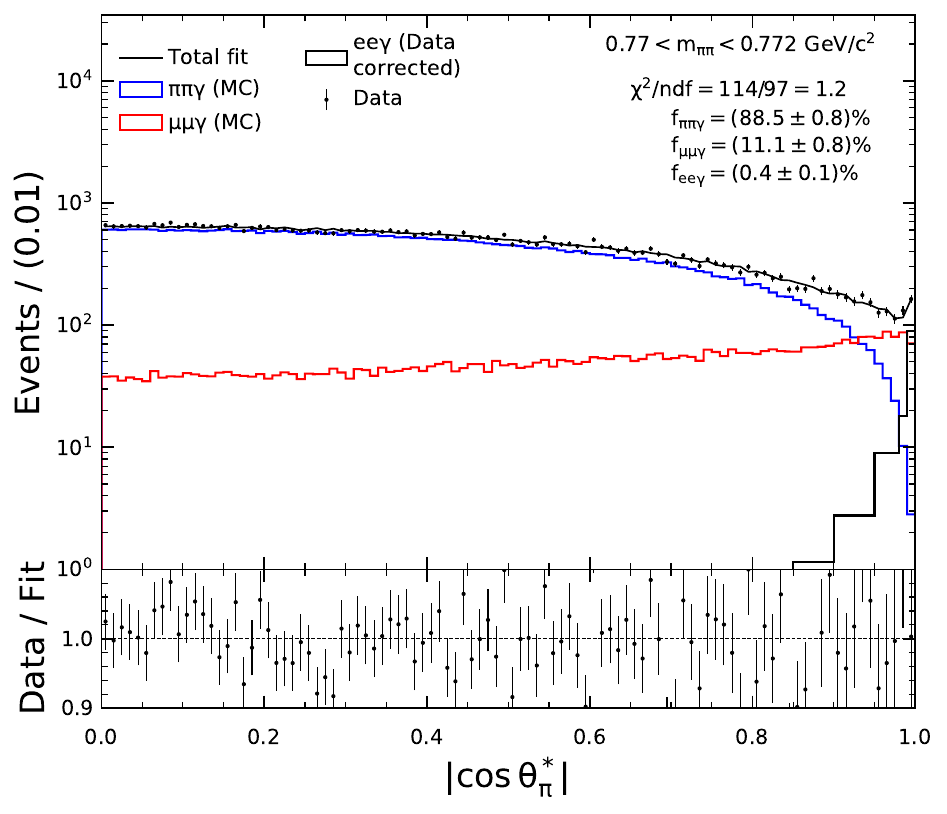}}
\vspace{-0.2cm}
\caption{
Angular fit results in two \mpipi\ bins, close to threshold (a) and at the $\rho$ peak (b). The quantities $f_{XX\gamma}\ (X=\pi,\mu,K,e)$ give the fitted fractions of the relevant processes (\KKg\ becomes negligible beyond 0.4\GeVM).}
\label{fig:fits}
\end{figure}

The fits to the data angular distributions consist in linear combinations of the normalized templates in more than 300 mass bins: 2\MeVM\ bins between $0.5-1$ \GeVM\ and 10\MeVM\ elsewhere, separately in both \mmumu\ and \mpipi\ masses to get the \mumug\ and \pipig\ spectra in their respective bases. Two fit results are illustrated in Figure~\ref{fig:fits}.

In order to reduce the sensitivity to systematic uncertainties due to the data-driven \eeg\ templates, a 3-step strategy is devised: a first fit is performed in the range $0.9<|\cos\theta^*|<1$ to obtain the normalization of \eeg\ events which peak in this region, followed by a second fit on \eeg-subtracted data in the lower range $0<|\cos\theta^*|<0.9$. The fitted \mumug\ and \pipig\ angular distributions are finally extrapolated up to $|\cos\theta^*|=1$ according to their templates to get the correct integral in each mass bin. Closure tests of accuracy of the fit on simulation have shown that resulting spectra are consistent with initial inputs.

Both \mumug\ and \pipig\ mass spectra are blinded until the last steps of the analysis, applying to each normalization a unique constant multiplicative factor. Trigger and tracking corrections to simulation are blinded in a similar way, bringing the total to six different blinding factors. These corrections, either on templates or mass-dependent, mitigate the data/MC efficiency differences from multiple sources, like the 2D-$\chi^2$ selection, for an overall negligible effect on the mass spectra shapes.

The dimuon data spectrum is compared to the QED prediction, evaluated as the \mumug\ MC spectrum corrected for ISR photon efficiency difference with data, shortcomings related to the overestimation of ``NLO"-type additional ISR and the absence of NNLO in \pho~\cite{isr}, and finally the imprecise description of vacuum polarization effects. The data/QED ratio being flat along the full \mmumu\ range, the \mumug\ data spectrum is unblinded and a constant fit to the ratio gives
\begin{equation}
    0.9955 \pm 0.0035_{\mathrm{stat}} \pm 0.0030_{\mathrm{syst}} \pm 0.0033_{\mathrm{\gamma ISR}} \pm 0.0043_{\mathrm{lumi}\ ee}\,\, .
\end{equation}
The first error comprises statistical uncertainties on the data spectrum (estimated following the bootstrap method~\cite{bootstrap}), on the QED prediction and on corrections; the second error includes systematic uncertainties from corrections and on the \eeg\ template determination; the last two errors, relevant only to this test, are due to the ISR photon efficiency correction and the uncertainty on the $e^+e^-$ luminosity. This shows a compatibility with unity within a precision of 0.71\% and validates the $\pi/\mu$ separation procedure.

The effective ISR luminosity, essential to the \pipig\ cross section measurement, is determined from the unfolded \mumug\ spectrum $dN_{\mu\mu}^\mathrm{ISR}/d\sqrt{s^\prime}$ as
\begin{equation}
\label{eq:lumi}
\frac{dL^\mathrm{eff}_\mathrm{ISR}}{d\sqrt{s^\prime}}=\frac{dN_{\mu\mu}^\mathrm{ISR}/d\sqrt{s^\prime}}{\epsilon_{\mu\mu}(\sqrt{s^\prime})\ \sigma^0_{\mu\mu}(\sqrt{s^\prime})}\,\, ,
\end{equation}
where $\epsilon_{\mu\mu}(\sqrt{s^\prime})$ is the acceptance (total efficiency) of the selection for this process and $\sigma^0_{\mu\mu}(\sqrt{s^\prime})$ is the bare \mumug\ cross-section without vacuum polarization. To smooth the large statistical fluctuations that affect the data spectrum, we replace the ratio $(dN_{\mu\mu}^\mathrm{ISR}/d\sqrt{s^\prime})/\epsilon_{\mu\mu}(\sqrt{s^\prime})$ with
\begin{equation}
\label{eq:replace}
\frac{dN_{\mu\mu}^\mathrm{MC\ gen}}{d\sqrt{s^\prime}}\times (1-f_\mathrm{LO\ FSR}) \times f_{\mu\mu}(\sqrt{s^\prime})\,\, ,
\end{equation}
that is the product of the dimuon \pho\ spectrum at generation level $dN_{\mu\mu}^\mathrm{MC\ gen}/d\sqrt{s^\prime}$, a factor $(1-f_\mathrm{LO\ FSR})$ that removes the LO FSR contribution and $f_{\mu\mu}(\sqrt{s^\prime})$, the fitted ratio of the unfolded data and MC \mumug\ spectra to a second-order polynomial function. This procedure is justified by the successful QED test and the fact that global corrections vary slowly over the full range.

The bare cross section for the \pipig\ final state is derived from the unfolded data spectrum as
\begin{equation}
    \sigma^0_{\pi\pi}(\sqrt{s^\prime})=\frac{dN_{\pi\pi}/d\sqrt{s^\prime}}{\epsilon_{\pi\pi}(\sqrt{s^\prime})\ dL^{\rm eff}_{\rm ISR}/d\sqrt{s^\prime}}\,\, ,
\end{equation}
where $\epsilon_{\pi\pi}(\sqrt{s^\prime})$ is the corresponding acceptance. Replacing the effective luminosity by its definition from Equations~\ref{eq:lumi} and \ref{eq:replace}, the bare cross section can be shown to be proportional to the ratio of the \pipig\ and \mumug\ spectra, which ensures the cancellation of common systematic effects from the ISR photon efficiency, $e^+e^-$ luminosity and vacuum polarization.

\begin{figure}[ht]
\centering
\subfloat[]{\includegraphics[width=0.48\columnwidth]{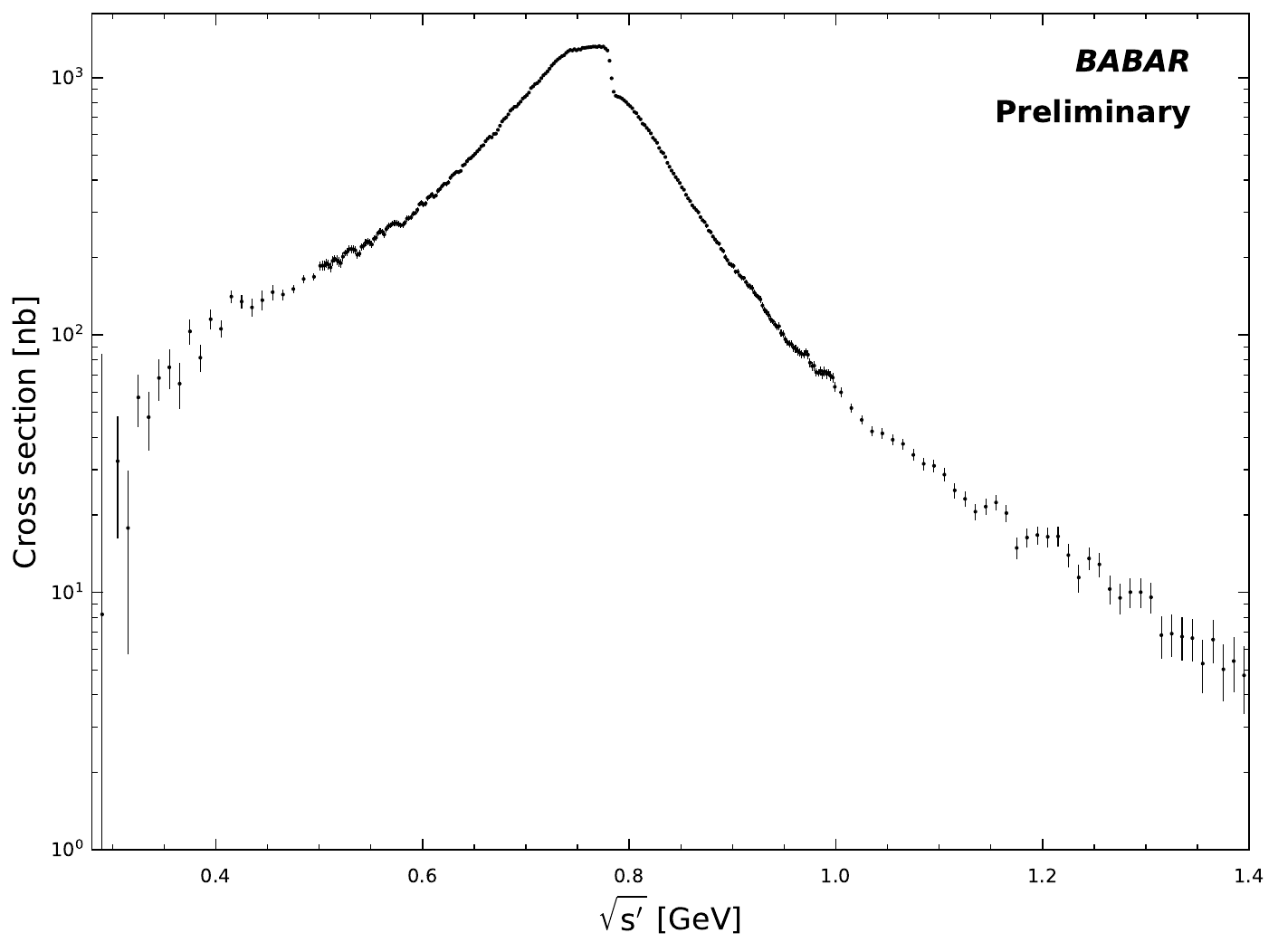}}
\hspace{0.4cm}
\subfloat[]{\includegraphics[width=0.48\columnwidth]{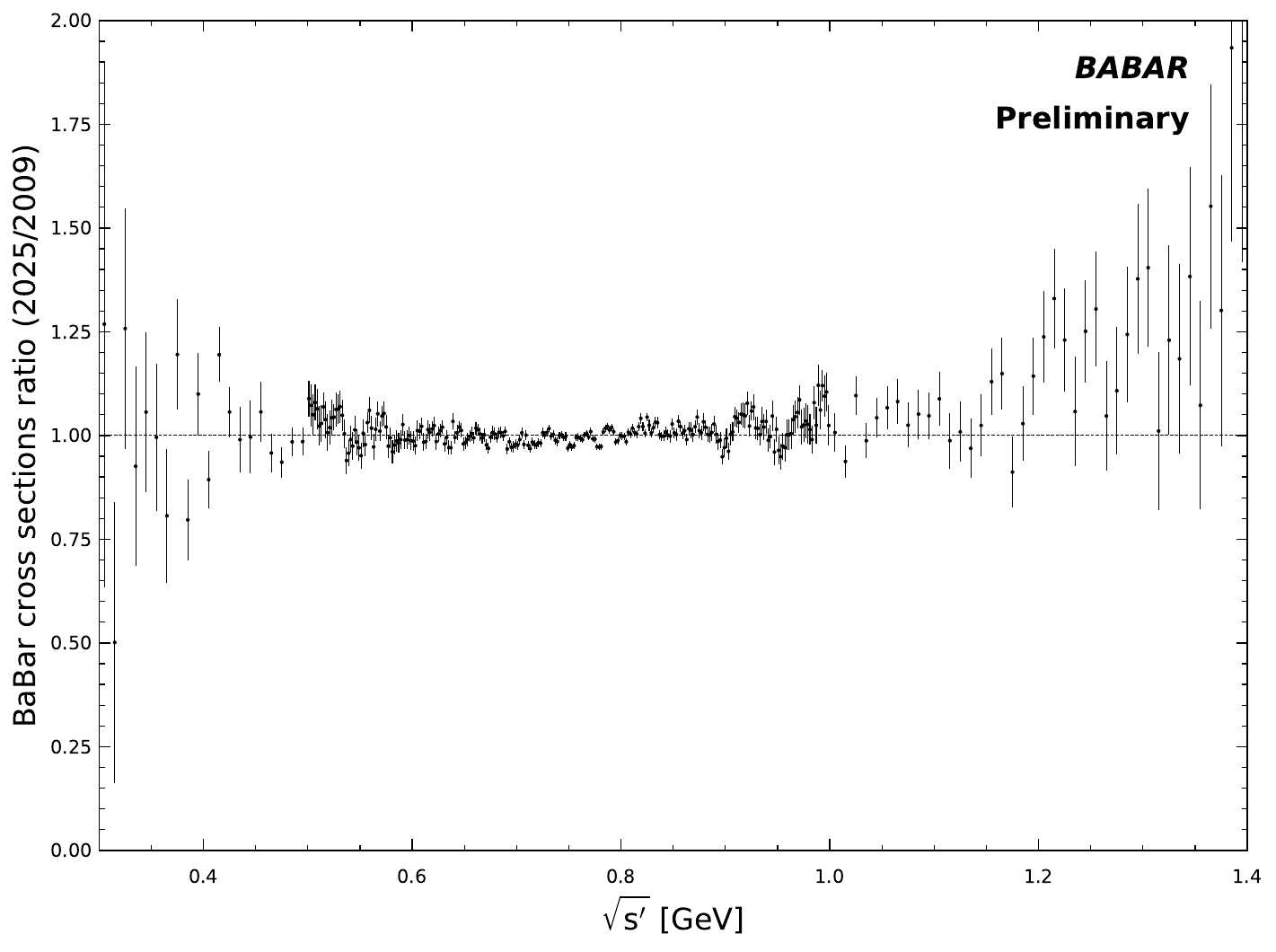}}
\vspace{-0.2cm}
\caption{
The \eepipig\ cross section measured in this work as a function of the reduced energy $\sqrt{s^\prime}$ (a) and the ratio of this measurement to the 2009 \babar\ result (b).}
\label{fig:cross_section}
\end{figure}

The measured \eepipig\ cross section and comparison to the 2009 result are presented in Figure~\ref{fig:cross_section} after unblinding. Both analyses appear to be consistent in most of the range from threshold to 1.4 GeV, except at large energies. The $\pi\pi$ contributions to $a_\mu$ below 0.5 GeV and between $0.5-1.4$ GeV are (in units of $10^{-10}$) $58.0 \pm 5.5 \pm 1.7$ and $456.2 \pm 2.2 \pm 1.7$, with statistical and systematic uncertainties, in excellent agreement with the respective results of the previous study $57.6 \pm 0.6 \pm 0.6$ and $455.6 \pm 2.1 \pm 2.6$. Because of the poor statistical separation of \pipig\ in regions where dimuon events dominate, the new measurements are not competitive at low energies, however they improve the systematic uncertainty in the higher range.

If combined from threshold to 1.8 GeV, the two studies lead to the value $(514.4 \pm 2.5)\times10^{-10}$, that is the most precise measurement of the $\pi\pi$ contribution to $a_\mu$ from a single experiment.

\section{Conclusion}

The contribution of the $\pi\pi$ channel to the anomalous magnetic moment of the muon, $a^{\pi\pi}_\mu$, is measured by \babar\ via ISR in a new blind analysis, using an independent method from the last 2009 measurement and twice as much data statistics, that is 460\fb. Final state separation in data is carried out with fits of angular distributions without relying on particle identification, previously the dominant source of systematic uncertainty. The effective ISR luminosity is obtained from the spectrum of the \eemumug\ process which is shown to be compatible with its QED prediction. In a preliminary result, the \eepipig\ cross section is found to be in good agreement with the 2009 measurement, the same conclusion applying to $a^{\pi\pi}_\mu$, here equal to $(58.0 \pm 5.5 \pm 1.7)\times10^{-10}$ below 0.5 GeV and $(456.2 \pm 2.2 \pm 1.7)\times10^{-10}$ between $0.5-1.4$ GeV, where the uncertainties are respectively statistical and systematic. This consistency proves the robustness of both analyses, which combined provide the most precise measurement of $a^{\pi\pi}_\mu$ from a single experiment.

\section*{Acknowledgments}

We are grateful for the extraordinary contributions of our PEP-II colleagues in achieving the excellent luminosity and machine conditions that have made this work possible. The success of this project also relies critically on the expertise and dedication of the computing organizations that support \babar, including GridKa, UVic HEP-RC, CC-IN2P3, and CERN. The collaborating institutions wish to thank SLAC for its support and the kind hospitality extended to them. We also wish to acknowledge the important contributions of J.~Dorfan and our deceased colleagues E.~Gabathuler, W.~Innes, D.W.G.S.~Leith, A.~Onuchin, G.~Piredda, and R. F.~Schwitters. This work benefited from funding by the French National Research Agency under contract ANR-22-CE31-0011 and from Laboratoire de physique nucléaire et de hautes énergies (UMR 7585, CNRS/IN2P3, Sorbonne Université, Université Paris-Cité).


\end{document}